\documentclass[%
 footinbib,
 twocolumn,
superscriptaddress,
longbibliography,
 amsmath,amssymb,
 aps,
 floatfix,
prb,
nobalancelastpage,
]{revtex4-1}

\usepackage{braket}
\usepackage{tikz}
\usepackage{dsfont}
\usepackage{siunitx, xspace}
\usepackage{array}
\usepackage{graphicx}
\usepackage{bbold}

\newcolumntype{M}[1]{>{\centering\arraybackslash}m{#1}}

\newcommand{\Hloc}{{H_{\text{loc}}}\xspace}
\newcommand{\Hbath}{{H_{\text{bath}}}\xspace}
\newcommand{\dt}{{\Delta t}\xspace}

\definecolor{OliveGreen}{rgb}{0.03, 0.47, 0.19}

\definecolor{boxgreen}{RGB}{15,157,88}
\definecolor{redtensor}{RGB}{219,68,55}
\definecolor{gateblue}{RGB}{66,133,244}

 \usepackage{pgfplots}
  \pgfplotsset{compat=newest}

\usetikzlibrary{external}
\usetikzlibrary{calc,  shapes, backgrounds, arrows, chains, matrix, positioning, scopes, decorations, decorations.pathmorphing,patterns, fit, pgfplots.groupplots, plotmarks}
\tikzset{>=latex}
\usetikzlibrary{external}
\usepackage{graphicx}
\usepackage{dcolumn}
\usepackage{bm}

\usepackage{hyperref}

\begin{document}

\title{Comparison of MPS based real time evolution algorithms for Anderson Impurity Models}

\author{Daniel Bauernfeind}
\email[]{daniel.bauernfeind@tugraz.at}
\affiliation{%
  Institute of Theoretical and Computational Physics\\
  Graz University of Technology, 8010 Graz, Austria
}%

\author{Markus Aichhorn}
\affiliation{%
  Institute of Theoretical and Computational Physics\\
  Graz University of Technology, 8010 Graz, Austria
}%

\author{Hans Gerd Evertz}
\email[]{evertz@tugraz.at}
\affiliation{%
  Institute of Theoretical and Computational Physics\\
  Graz University of Technology, 8010 Graz, Austria
}%

\date{\today}

\begin{abstract}
We perform a detailed comparison of two Matrix Product States (MPS) based time evolution algorithms for Anderson 
Impurity Models. To describe the bath, we use both the star-geometry as well as the commonly employed Wilson chain 
geometry. For each bath geometry, we use either the Time Dependent Variational Principle (TDVP) or the Time Evolving Block 
Decimation (TEBD) to perform the time evolution. To apply TEBD for the star-geometry, we use a specially adapted algorithm that can 
deal with the long-range coupling terms. Analyzing the major sources of errors, one expects them to be 
proportional to the system size for all algorithms. Surprisingly, we find errors independent of system size except for 
TEBD in chain geometry. Additionally, we show that the right combination of bath representation and time evolution 
algorithm is important. While TDVP in chain geometry is a very precise approach, TEBD in star geometry is much faster, 
such that for a given accuracy it is superior to TDVP in chain geometry. This makes the  adapted version of TEBD in star geometry the most 
efficient method to solve impurity problems. 
\end{abstract}

\maketitle

\section{Introduction}
Describing strongly correlated electron materials is among the most difficult tasks in solid state physics. One 
breakthrough in this field was the development of the Dynamical Mean Field Theory 
(DMFT)~\cite{MetznerVollhardt_Dinf,Georges_DMFToriginal,GeorgesDMFT}. DMFT accounts for local electronic correlations by a 
self-consistent mapping of a lattice problem, describing the low-energy subspace of the material, onto an Anderson 
Impurity Model (AIM)~\cite{AndersonOrig}. The subsequent solution of this impurity problem is the most important part 
of a DMFT calculation. At present, Continuous Time 
Quantum Monte Carlo (CTQMC)~\cite{WernerFirstCTQMC,GullCTQMC} is the work-horse method for DMFT real material 
calculations. 
Many other approaches exist, like the Numerical Renormalization Group~\cite{WilsonNRG,BullaNRG}, Configuration 
Interaction~\cite{HaverkortED,ZgidCI,Zaera_CI}, and also methods based on the Density Matrix Renormalization Group 
(DMRG) and the related Matrix Product States (MPS)~\cite{WhiteDMRG, SchollwoeckDMRG_MPS}.
\\
One of the major advantages of MPS based impurity solvers is that they can give access to real-frequency 
spectral functions by employing real-time evolution. 
While these methods provide excellent results for the single orbital case, adding more and more orbitals is a very 
challenging task. 
To overcome this issue, the Fork Tensor Product States (FTPS)~\cite{FTPS} solver was recently developed and applied to several materials~\cite{FTPS,FTPS_SMO}. This new 
approach has allowed to resolve multiplets in the Hubbard bands of the single particle excitation spectrum for real 
materials, which have also been observed experimentally~\cite{FTPS_SMO} but are inaccessible from the imaginary time 
results from standard CTQMC methods. It is therefore important to identify the most efficient methods for real-time 
evolution of impurity models.

A recent review~\cite{Schollwoeck_CompareTevoAlgs} reported an extensive comparison of 
several MPS-based time evolution algorithms for different types of models. It did not, however, include impurity 
problems. They are special in the sense that most of the degrees of freedom are non-interacting. 
The conclusion of Ref.~\onlinecite{Schollwoeck_CompareTevoAlgs} was that methods perform better or worse 
depending on the 
model or observable studied. In all cases, the Time Dependent Variational Principle 
(TDVP)~\cite{HaegemanTDVP1,HaegemanTDVP2} was among the most 
reliable approaches, while the global Krylov method although accurate tended to be too time-consuming. 

Using tensor networks, impurity problems usually have been transformed to the 
Wilson chain-geometry representation of the bath - essentially a nearest neighbor tight-binding chain. 
The so-called star geometry on the other hand involves hopping processes from the impurity to all bath sites that 
become long ranged 
when mapped onto a linear chain. Although the reformulation of DMRG as a variational principle on the space of MPS 
allows to deal with such long-range terms, the chain geometry was believed to be the superior representation of 
impurity models.
Surprisingly, Wolf \emph{et al.}~\cite{WolfStarG} demonstrated that for MPS, the star geometry is in fact a more 
economic representation. 
Wolf \emph{et al.} used the global Krylov method to perform the time evolution which can be, as mentioned above, very 
time-consuming. As an alternative, some of us published a modified Time Evolving Block Decimation (TEBD) 
~\cite{VidalTEBD1,VidalTEBD2,DaleySchollwoek_tDMRG} algorithm for the star geometry in Ref.~\onlinecite{FTPS}. 
There is also the possibility to use the TDVP, which has been argued to be suited to long-range terms ~\cite{HaegemanTDVP1}. 
Therefore, open questions remain regarding the time evolution methods used for impurity 
problems, for example: Does the bath geometry affect the accuracy of the result? If so, does this depend on the 
time evolution algorithm? And arguably the most important one: Which method is best?

In the present paper, we use TDVP and TEBD as time evolution algorithms for AIMs in the star and chain geometry and 
perform an in-depth comparison of the four possibilities.
We compare the quantity of interest in DMFT calculations, i.e., the impurity Green's function.
We show that to obtain precise results, the correct combination of bath representation and time evolution algorithm is 
crucial. Specifically we demonstrate that the adapted TEBD~\cite{VidalTEBD1,VidalTEBD2,DaleySchollwoek_tDMRG} is more 
accurate in the star geometry, while TDVP gives better results in the chain geometry. 
We find that the TEBD approach in star-geometry is much faster than TDVP in the chain geometry when a specific precision of the Green's function is prescribed. 
Additionally, we focus on the behavior of the algorithms as a function of bath size. Surprisingly, all algorithms except 
TEBD in chain-geometry have errors nearly independent of system size. 
This very favorable scaling is especially important for DMFT calculations, as one wants to use a large number of sites 
to represent the bath hybridization well.

This paper is structured as follows. In Sec.~\ref{sec:impmodel} we introduce AIMs in star-geometry and in 
chain-geometry representation of the bath and discuss how to obtain bath parameters from the bath hybridization.  
In Sec.~\ref{sec:MPS}, we briefly introduce MPS and how to obtain Green's functions using real-time evolution. In 
Sec.~\ref{sec:TevoAlgs} we discuss the different time evolution methods, including a discussion of the various 
error sources. Sec.~\ref{sec:results_U0} contains results, first for the non-interacting case, and in 
Sec.~\ref{sec:results_U} for the interacting case. Finally, Sec.~\ref{sec:conclusions} contains the conclusions.

\section{Anderson Impurity Models}\label{sec:impmodel}
In the star geometry, an impurity described by a local interacting Hamilton $\Hloc$ is coupled to a bath of 
free fermions via hopping terms from the impurity to \emph{every} bath site. For a one-band model this results in:
\begin{align} \label{eq:AIM_Star}
  H_{\text{star}} &= H_{\text{loc}} + H_{\text{bath}}   \\
  H_{\text{loc}} &= U n_{I,\uparrow} n_{I,\downarrow} + \epsilon_I \left(n_{I,\uparrow} + n_{I,\downarrow} \right) \nonumber\\
  H_{\text{bath}} &= \sum_{k,\sigma} H_{k,\sigma}=  \sum_{k,\sigma} \epsilon_k n_{k,\sigma} \nonumber + V_k \left( c^\dag_{I,\sigma} c_{k,\sigma} + c^\dag_{k,\sigma} c_{I,\sigma} \right). \nonumber
\end{align}
Here, $c_{k,\sigma}^\dag$ $(c_{k,\sigma})$ create (annihilate) an electron at bath site $k$ with spin $\sigma$, 
$n_{k,\sigma} =c^\dag_{k,\sigma} c_{k,\sigma}$ are the usual particle number operators and we label the impurity degrees 
of freedom by an index $I$. If one puts the sites of $H_{\text{star}}$ on a one-dimensional manifold (like an MPS - see 
below), the hopping terms become long-range.
\\~\\
Using a Lanzcos-like tridiagonalization, one can map the star geometry onto the so-called Wilson 
chain~\cite{WilsonNRG,BullaNRG} or just chain-geometry, where the impurity couples to the first bath site only: 
\begin{align}\label{eq:AIM_Chain}
  H_{\text{chain}} &= \Hloc + t_0\left(c^\dag_{I,\sigma} c_{1,\sigma}+ c^\dag_{1,\sigma} c_{I,\sigma} \right) +\\
  &\sum_{i, \sigma} t_i\left(c^\dag_{i,\sigma} c_{i+1,\sigma}+ c^\dag_{i+1,\sigma} c_{i,\sigma} \right) + \bar{ \epsilon }_i n_{i\sigma}.\nonumber
\end{align}

\subsection{Determination of Bath parameters}
The bath can also be described by the continuous (DMFT~\cite{GeorgesDMFT}) hybridization function $\Delta(\omega)$.
By tracing out the bath in Eq.~\eqref{eq:AIM_Star}, one finds:
\begin{subequations}
\begin{align}\label{eq:bathHyb}
  \Delta(\omega) &= \sum_k \frac{V_k^2}{\omega - \epsilon_k +i0^+} \\
  -\frac{1}{\pi} \mathcal{I} \Delta(\omega) &\approx \sum_{k=1}^N V_k^2 \frac{\eta}{\pi \left(  \left(\omega-\epsilon_k 
\right)^2 + \eta^2 \right) }. \label{eq:Discr_approx}
\end{align}
\end{subequations}
A continuous bath corresponds to an infinite sum. In the last line, we used only a finite number of $N$ bath sites, and 
approximated the delta-peaks by Lorentzians of finite width $\eta$. 
Usually, the hybridization function $\Delta(\omega)$ is given and in order to map it to an AIM of finite size, one 
needs to find values $\epsilon_k$ and $V_k$, such that Eq.~\eqref{eq:Discr_approx} is as good approximation.
In the present study, we choose to split the $\omega$-axis into $N$ equidistant intervals $I_k$ of size $\Delta 
\epsilon$. We describe each interval using a single bath site with on-site energy $\epsilon_k = \min I_k + \frac{\Delta 
\epsilon}{2}$, where $\min I_k$ is the minimum of interval $I_k$. The hopping amplitude $V_k$ can then be computed 
from~\cite{BullaNRG, WolfStarG}: 
\begin{equation}
  V_k^2 = \int_{I_k} -\frac{1}{\pi} \mathcal{I} \Delta(\omega).
\end{equation}
A subsequent basis transformation into the Lanzcos-basis of $\Hbath$ yields the bath parameters in the chain 
geometry~\cite{BullaNRG, WolfStarG}. Note that for particle hole symmetry, the on-site energies $\bar{ \epsilon }_i$ of 
the chain geometry are exactly zero~\cite{BullaNRG}. For both geometries, we enforce particle hole symmetry in the bath 
parameters. In the following, we will give all results in units of the half bandwidth of the bath spectral function 
$-\frac{1}{\pi} \mathcal{I} \Delta(\omega)$.


\section{Matrix Product States}\label{sec:MPS}

\begin{figure}
  \centering
  \includegraphics{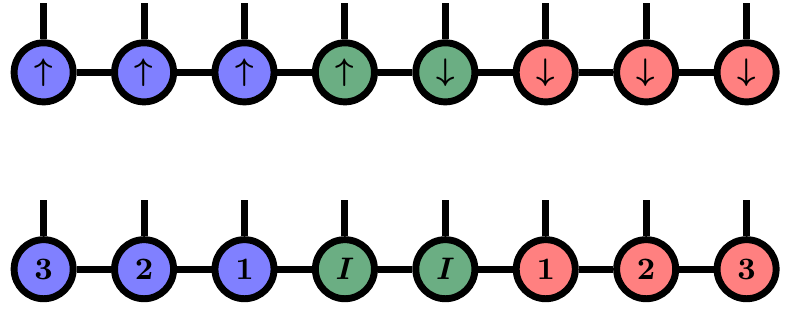}
  \caption{Graphical representation of an MPS used for an AIM with $N=3$ bath sites. \emph{Top:} To separate the spin 
species, we place the impurity in the middle (green circles) and attach the spin-up (spin-down) bath to the left (right) 
colored in blue (red). \emph{Bottom:} Labeling of sites in the star- as well as in the chain-geometry. The bath sites are 
arranged in ascending order with increasing distance from the impurity.}
 \label{fig:MPS_ImpModel}
 \end{figure}

MPS are an efficient parametrization of quantum mechanical states as a product of local matrices. Consider a system 
consisting of $2N+2$ sites with a local basis $\ket{s_i}$ at site $i$:
\begin{equation}
  \ket{\psi} = \sum_{ \{s\} } c_{s_1, \cdots s_{2N+2} } \ket{ s_1 \cdots s_{2N+2} }.
\end{equation}
In an MPS, the coefficient $c_{s_1, \cdots s_{2N+2} }$ is factorized into a product of matrices using repeated Singular 
Value Decompositions (SVDs)~\cite{SchollwoeckDMRG_MPS}:
\begin{equation}
  c_{s_1, \cdots s_{2N+2} } =  A^{s_1} \cdot A^{s_2} \cdots A^{s_{2N+1}}.
\end{equation}
Each $A_i^{s_i}$ is a rank-$3$ tensor, except the two tensors at the edges which are of rank-$2$. Since $s_i$ represents 
the local basis $\ket{s_i}$, this index is called physical. The \emph{matrix} indices which are summed over are called 
bond-indices. The dimension of the bond indices (bond dimension) is the number of Schmidt values kept during the 
calculation, implying that some Schmidt values are discarded. The sum of the square of all discarded Schmidt values is called 
truncated weight $t_w$~\cite{SchollwoeckDMRG_MPS}. MPS and other tensor networks are often depicted using a 
graphical representation as in Fig.~\ref{fig:MPS_ImpModel}. 
Since the Hamiltonian of an AIM connects the two spin species only via interactions on the impurity (see 
Eq.~\eqref{eq:AIM_Star}), it has turned out to be favorable to separate them in the MPS, and to use a local Hilbert 
space 
of dimension two (empty and occupied) for each site~\cite{Saberi_FoldedVSunfolded,Ganahl2BHubb, FTPS}. Thus, we place 
the 
impurity in the middle of the chain and connect the spin-up (spin-down) bath to its left (right), as shown in 
Fig.~\ref{fig:MPS_ImpModel}.
\\
Using this arrangement of sites, we calculate the ground state $\ket{\psi_0}$ and its energy $E_0$ using DMRG. The 
central object of interest in DMFT calculations is the Green's function of the impurity. In the following, we will hence
focus our attention on the greater Green's function of the impurity (omitting the spin index):
\begin{align}\label{eq:GreaterGF}
  G^> (t) &= \bra{ \psi_0 } c_I e^{-i H t} c_I^\dag \ket{ \psi_0 } e^{iE_0t} \\
  &=  \left( e^{i H \frac{t}{2}} c_I^\dag\ket{ \psi_0 } \right)^\dag \left( e^{-i H \frac{t}{2}} c_I^\dag \ket{ \psi_0 } 
\right ) e^{iE_0t} .\nonumber
\end{align}
To calculate Eq.~\eqref{eq:GreaterGF}, we first apply the operator $c^\dag_I$ onto the ground state. Next, as indicated 
by brackets in the the second line of Eq.~\eqref{eq:GreaterGF}, we perform two separate time evolutions up to time 
$\frac{t}{2}$~\cite{Ganahl2BHubb,Karasch_ExtendingRange} and calculate the overlap. We checked that the results 
are the same for the lesser Green's function and also checked that non-particle hole symmetric models show a behavior very similar to the results presented below.


\section{Time Evolution Algorithms} \label{sec:TevoAlgs}
In the following, we discuss the different time evolution algorithms with special attention to our adapted TEBD approach in star 
geometry, and we analyze the main sources of error.
\subsection{TEBD}
Several slightly different formulations of 
TEBD~\cite{VidalTEBD1,VidalTEBD2,DaleySchollwoek_tDMRG,Schollwoeck_CompareTevoAlgs} exist. The common strategy is 
to split the full time evolution operator of some small time step $\tau$ into manageable parts which can be applied to 
evolve the state forward in time.
Specifically, one employs Suzuki-Trotter breakups~\cite{SuzukiDecomp} to obtain a decomposition into operators 
acting on two sites only. The action of such a \emph{gate} merges the MPS tensors of the two sites, which are 
then separated again using a SVD combined with a truncation.
Apart from the truncation, the main error of this approach is caused by the (in our case second order) Trotter breakup:
\begin{align} \label{eq:secondorderbreakup}
  e^{ \tau (A+B) } &= e^{-i A \frac{\tau}{2} }  e^{-iB\tau}e^{-iA \frac{\tau}{2}} + \frac{\tau^3}{12} \mathcal{C} + 
\mathcal{O}(\tau^4) \\
  \mathcal{C}  &= \left( \frac{1}{2} \big[ A,[A,B] \big]+\big[B,[A,B]\big] \right) \nonumber.
\end{align}
According to Eq.~\eqref{eq:secondorderbreakup}, the total error of the time evolution is not only determined by the 
time step $\tau$, but also by the matrix elements of the double commutators $\mathcal{C}$. Since the latter are 
different for the star and the chain 
geometry, the error of TEBD in both geometries can be very different. This will be discussed in the following two 
subsections.

\subsubsection*{Chain Geometry} \label{sec:ChainGeometry}
To employ TEBD in the chain-geometry, we use the standard second order breakup between even and odd 
terms~\cite{SchollwoeckDMRG_MPS}. We write the Hamiltonian of the chain geometry $H_{\text{chain}} = \sum_i h_{i,i+1}$ 
as a sum of local terms $h_{i,i+1}$ acting on nearest neighbor sites $i$ and $i+1$ only and define:
\begin{align}\label{eq:define_evenodd}
  H_{\text{chain}} &= H_{\text{even}} + H_{\text{odd}} \nonumber \\
  H_{\text{even}} &= \sum_{i:\text{even}}h_{i,i+1} \text{, } \nonumber\\
  H_{\text{odd}} &= \sum_{i:\text{odd}}h_{i,i+1} \nonumber\\
  e^{-i H_{\text{chain}} \dt  } &\approx e^{-i H_{\text{odd}} \frac{\dt}{2}}  e^{-iH_{\text{even}} \dt} e^{-i 
H_{\text{odd}} \frac{\dt}{2}}. 
\end{align} 
In the particle hole symmetric case, the on-site energies $\bar{\epsilon_i}$ are exactly zero, removing any ambiguity 
of how to distribute the on-site terms among even- and odd parts of the Hamiltonian.
For particle hole symmetry ($\bar{\epsilon_i}=0$), we evaluated the double commutators of 
Eq.~\eqref{eq:secondorderbreakup}. They turn out to be various sums over hopping terms multiplied by three amplitudes:
\begin{equation}\label{eq:TrotterErrChain}
  \mathcal{C}_{\text{chain}} = -\frac{1}{2} \sum_{i:\text{even}} t_i t_{i+1} t_{i+2} \left( c^\dag_i c_{i+3} + 
c^\dag_{i+3} 
c_i\right)  + \cdots.
\end{equation}
Additionally, the interaction $U$ gives two terms that couple to the neighbors of the impurity, independent of $N$. For 
the scaling with bath size the latter terms can be neglected, since they are of order $\mathcal{O}(1)$. Each of the 
individual sums of Eq.~\ref{eq:TrotterErrChain} scales linearly with system size $N$, 
they contain $N$ terms of order $1$, and the hopping amplitudes $t_i$ are independent of 
$N$~\footnote{For example a semi-circular bath spectral function can be represented by $t_i = t$ for all sites. Note 
that we use the Lanzcos tri-diagonalization instead.}. Thus, one should expect the overall error of TEBD 
in chain geometry to scale like the bath size $N$.

\subsubsection*{Star Geometry}

\begin{figure}
  \centering
  \includegraphics{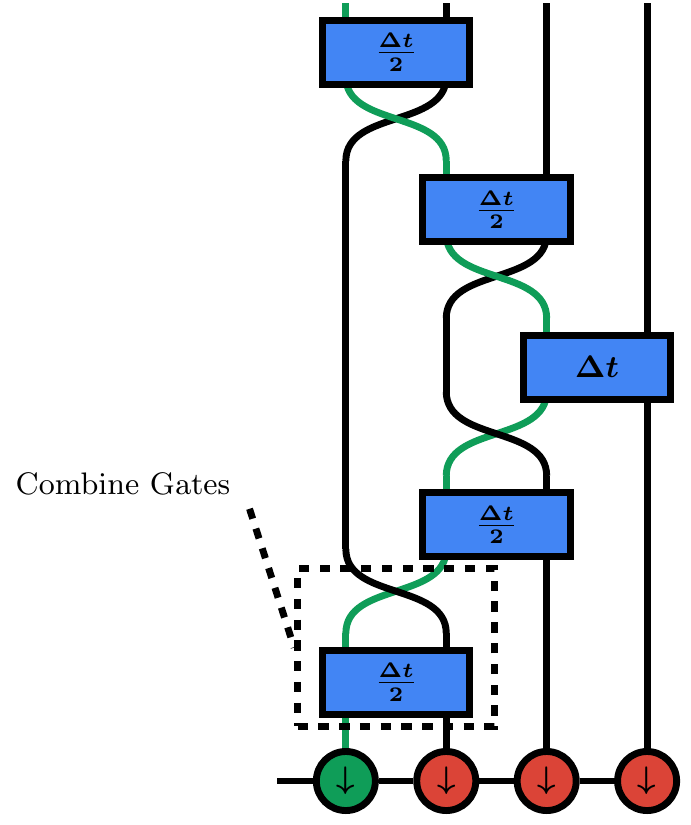}
  \caption{Application of the time evolution operator of Eq.~\eqref{eq:iterativeDecomp} onto the spin-down bath. We use 
swap gates depicted by two crossing arcs to evolve the long-range hopping terms. The green line visualizes the 
position of the impurity as it is moved through the bath during the time evolution. All gates except the one with 
$H_{N,\sigma}$ are calculated with a time step $\frac{\dt}{2}$, reflecting the use of the second order breakup. Note 
that with this approach one can combine every swap gate with an actual time evolution gate, thus avoiding any 
additional SVDs. 
}
 \label{fig:TevoSwapGates}
 \end{figure}

To perform the TEBD time evolution of an AIM in star geometry (Eq~\eqref{eq:AIM_Star}), we employ an approach we recently 
developed (see Refs~\onlinecite{FTPS, THESIS}), which is based on swap-gates~\cite{OrusSwap1,VerstraeteSwap2} and iterative Suzuki-Trotter decompositions. 
First, we first split off $\Hloc$ from the time evolution operator for some small time-step $\dt$ in a second order 
Trotter decomposition:
\begin{equation}\label{eq:StarTrotterHloc}
  e^{-iH\dt} \approx e^{-i \Hloc \frac{\dt}{2}} \left( \prod_{\sigma} e^{-i\sum_k H_{k,\sigma} \dt} \right ) e^{-i 
\Hloc 
 \frac{\dt}{2}},
\end{equation}
where $H_{k,\sigma}$ was defined in Eq.~\eqref{eq:AIM_Star}. Then we split off the term containing $H_{1,\sigma}$, 
then $H_{2,\sigma}$ and so on, until we obtain:
\begin{equation}\label{eq:iterativeDecomp}
  e^{-i\sum_k H_{k,\sigma} \dt} \approx \left( \prod_{k=1}^{N} e^{-i H_{k,\sigma}  \frac{\dt}{2}} \right) \left( 
\prod_{k=N}^1  e^{-i H_{k,\sigma} \frac{\dt}{2}} \right).
\end{equation}
The Trotter errors of these approximations will be discussed below.
Note that Eq.~\eqref{eq:iterativeDecomp} is a product of operators acting on two sites only, but all terms except 
$H_{1,\sigma}$ involve sites that are not nearest neighbors in the MPS. To 
be able to apply such non-nearest neighbor gates, we use swap-gates. Their purpose is to swap the degrees of freedom of 
two sites in the MPS-tensor network. For a fermionic Hilbert space with local dimension of $2$ (empty $\ket{0}$ or 
occupied $\ket{1}$), a swap-gate $S$ is given by:
\begin{equation}
  S = \ket{00}\bra{00} + \ket{01}\bra{10} + \ket{10}\bra{01} - \ket{11}\bra{11}.
\end{equation}
Note the minus sign in the last term, resulting from the exchange of two fermions~\cite{THESIS}. The process of applying 
all the gates in Eq.~\eqref{eq:iterativeDecomp} is depicted in Fig.~\ref{fig:TevoSwapGates} and described in the 
following.\\

Instead of a gate with $H_{1,\sigma}$ only, we apply a combined gate $S \cdot e^{-i H_{1,\sigma}\frac{\dt}{2} }$, 
meaning that we first time evolve and swap afterwards. The impurity degrees of freedom are then located at what 
was previously the first bath site and is now a nearest neighbor of the second bath site. We continue by applying a 
combined two-site gate $S \cdot e^{-i H_{2,\sigma} \frac{\dt}{2}}$ after which the impurity and bath site $3$ are 
nearest neighbors and so on. When the impurity arrives at site $N-1$, we time evolve with $H_{N,\sigma}$ (without 
swapping). Since we used a second order decomposition, we have to re-apply all gates with $k=N-1 \cdots 1$. This time, 
though, we have to swap first and time evolve afterwards, since otherwise we would have to take care of an additional 
fermionic sign in the hopping terms~\cite{THESIS}. This means that we apply a two-site gate $e^{-i 
H_{N-1,\sigma}\frac{\dt}{2} } \cdot S$ followed by $e^{-i H_{N-2,\sigma}\frac{\dt}{2} } \cdot S$, etc., until the 
impurity is back in the middle of the MPS and every term in Eq~\eqref{eq:iterativeDecomp} is dealt with.
\\ ~ \\
Now let us evaluate the Trotter errors due to these various decompositions. In the first breakup, we split off $\Hloc$ 
from the 
rest of the Hamiltonian, i.e., $A=\Hloc$ and $B=\Hbath$ in Eq.~\eqref{eq:secondorderbreakup}. Evaluation of the 
double commutators yields:
\begin{align}\label{eq:TrotterErr_HlocStar}
  &\mathcal{C}_{\Hloc} = -2\sum_\sigma K_\sigma n_{I,\sigma}  \sum_k V_k^2 \\
  &+ \sum_{k,\sigma} V_k \left( c^\dag_{I,\sigma} c_{k,\sigma} + c^\dag_{k,\sigma} c_{I,\sigma} \right ) K_\sigma \left( 
\frac{1}{2} K_\sigma - \epsilon_k \right) \nonumber \\
  &+ \sum_{k,k',\sigma} V_k V_{k'}  K_\sigma \left( c^\dag_{k,\sigma} c_{k',\sigma} + c^\dag_{k',\sigma} c_{k,\sigma} 
\right) \nonumber \\ 
  &- \sum_{k,k',\sigma} V_k V_{k'} U \left( c^\dag_{I,\sigma} c_{k',\sigma} - c^\dag_{k',\sigma} c_{I,\sigma} \right)  
\left( c^\dag_{I,\bar{\sigma}} c_{k,\bar{\sigma}} - c^\dag_{k,\bar{\sigma}} c_{I,\bar{\sigma}} \right)  \nonumber, \\
  ~ \nonumber
\end{align}
with $K_\sigma = U n_{I\bar{\sigma}}+\epsilon_0$ and $\bar{\sigma}$ the spin in opposite direction of spin $\sigma$.
\\
Next, we split off $H_{1,\sigma}$ from $\sum_{k>1} H_{k,\sigma}$, i.e., $A=H_{1,\sigma}$ and $B=\sum_{k>1} 
H_{k,\sigma}$. The error of this decomposition  (with $j=1$):
\begin{align}\label{eq:TrotterErr_HbathStar}
  \mathcal{C}_{H_{j,\sigma}} &= \frac{1}{2} \sum_{k>j} V_j^2 V_k \left( c^\dag_{I,\sigma} c_{k,\sigma} + 
c^\dag_{k,\sigma} c_{I,\sigma} \right ) \\
  &- V_j \left( c^\dag_{I,\sigma} c_{j,\sigma} + c^\dag_{j,\sigma} c_{I,\sigma} \right ) \sum_{k>j} V_k^2 \nonumber \\
  &+ \sum_{k>j} V_j V_k \left( c^\dag_{j,\sigma} c_{k,\sigma} + c^\dag_{k,\sigma} c_{j,\sigma} \right ) \left( 
\frac{1}{2} \epsilon_j - \epsilon_k \right) \nonumber.
\end{align}
The next decomposition separating $H_{2,\sigma}$ from $\sum_{k>2} H_{k,\sigma}$ has the exact same form as above now with $j=2$, and 
hence an error of $\mathcal{C}_{ H_{j,\sigma}}$ for $j=2$. Iterating over all decompositions, the total error of the breakup used to separate the 
time evolution operator in the star geometry becomes:
\begin{equation}\label{eq:TrotterErr_StarTot}
  \mathcal{C}_{\text{star}} = \mathcal{C}_\Hloc + \sum_{j\neq N, \sigma} \mathcal{C}_{ H_{j,\sigma} }.
\end{equation}
Let us analyze how this error scales with the number of bath sites $N$. From Eq.~\eqref{eq:bathHyb}, we find that $V_k 
\sim \frac{1}{\sqrt{N}}$. Hence, $\mathcal{C}_\Hloc$ in Eq.~\eqref{eq:TrotterErr_HlocStar} scales linearly with $N$, since 
each 
of the last two lines contains $N^2$ terms of order $1$ multiplied by $V_k^2 \sim \frac{1}{N}$. \\
Similarly, the other terms in Eq.\ref{eq:TrotterErr_StarTot} give a growth of the error not faster than $N$ (two terms in $\mathcal{C}_\Hloc$ also have $\sim N$ scaling and the next lowest order is a $\sqrt{N}$-scaling). We hence should expect the error in the star geometry to scale linearly with $N$.
\\
This might seem surprising at first, since to obtain Eq.~\eqref{eq:iterativeDecomp}, we performed $N$ individual 
decompositions and one might therefore expect the total error to scale at least $\sim N^2$. However, since the hopping 
parameters are $N$-dependent themselves ($V_k \sim \frac{1}{\sqrt{N}}$), the overall error improves by a factor of 
$\frac{1}{N}$. Note that the actual errors in the calculations will depend on the matrix elements of the error terms Eq.\eqref{eq:TrotterErrChain} or Eq.~\eqref{eq:TrotterErr_StarTot}.

\subsection{TDVP}
Instead of approximating the time evolution operator, TDVP directly solves the time dependent Schr\"odinger equation, 
albeit only in a restricted subspace - in the space of MPS of fixed bond dimension. TDVP constructs a projection 
operator $P_{ T_{\ket{\psi}} }$ that projects the right hand side of the Schr\"odinger equation onto the 
tangent space $T_{\ket{\psi}}$ of the current MPS $\ket{\psi}$:

\begin{equation}
 \frac{\partial}{\partial t} \ket{\psi} = -i P_{ T_{\ket{\psi}} } H \ket{\psi}.
\end{equation}

This equation then results in a set of equations that can be integrated with an approach very similar to DMRG, 
replacing the ground-state search by Krylov time propagation~\cite{HaegemanTDVP2, Schollwoeck_CompareTevoAlgs}. 
In the present publication we employ two-site TDVP (2TDVP in Ref.~\onlinecite{Schollwoeck_CompareTevoAlgs}) using the 
second order integrator by sweeping left-right-left with half time step $\frac{\Delta t}{2}$. For additional details on 
TDVP we refer to the existing literature~\cite{HaegemanTDVP1,HaegemanTDVP2,Schollwoeck_CompareTevoAlgs}. Apart from 
MPS-matrix truncation 
and finite time step $\Delta t$, TDVP has an additional parameter, namely when to terminate the Krylov series. 
We stop creating new Krylov vectors when the total contribution of two consecutive vectors to the matrix 
exponential is less than $10^{-15}$.

The errors of TDVP are, apart from the truncation, a time step error similar to TEBD of order $\Delta t^3$ and an error 
due 
to the projection of the Schr\"odinger equation~\cite{Schollwoeck_CompareTevoAlgs}. We expect the latter to strongly 
depend on the bath representation, because it is exactly zero for Hamiltonians with nearest neighbor terms 
only~\cite{Schollwoeck_CompareTevoAlgs,HaegemanTDVP2}. As a function of bath size, we expect an error linear in 
$N$, since TDVP approximates $N$ coupled equations by integrating them one after the other.

\section{Results}\label{sec:results_U0}

\begin{figure}
  \centering
  \includegraphics{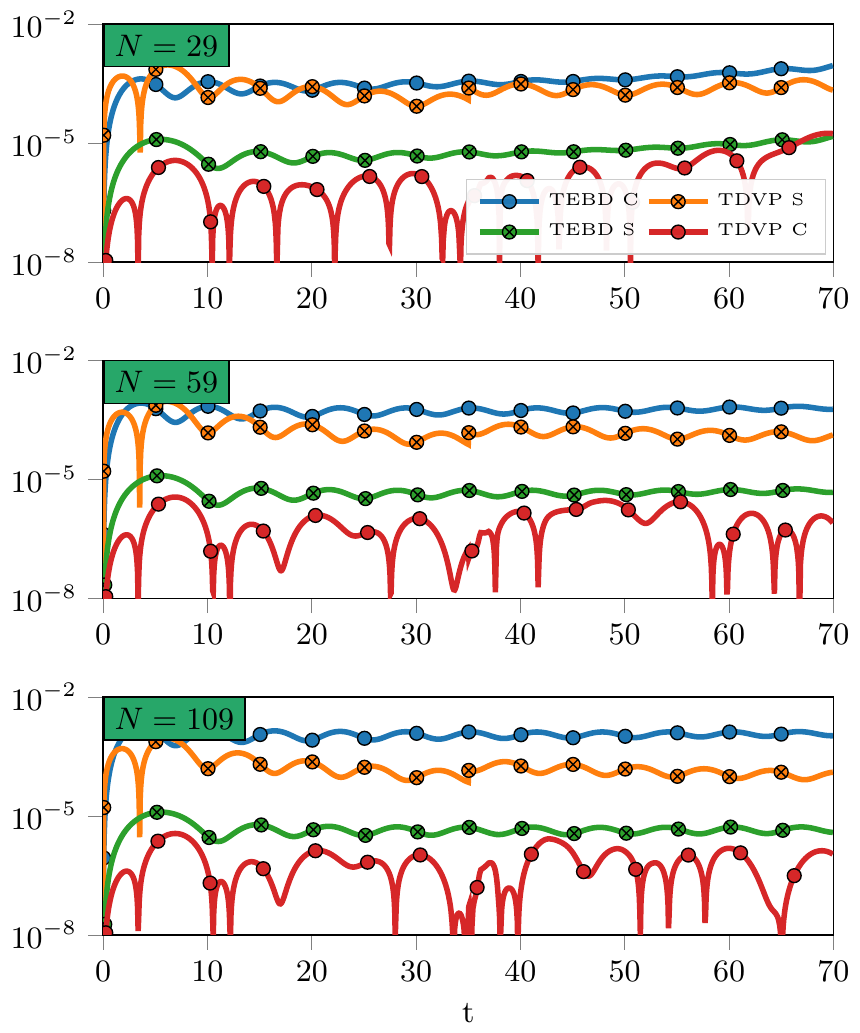}
  \caption{Absolute differences $ | \Re G^{>}_{\text{exact}}(t) - \Re G^{>}_{\text{DMRG}}(t)|$ for chain- (C) and star 
(S) geometries and various bath sizes ($N = 29, 59 ,109$ from top to bottom) at $U=0$. The bath for this calculation 
was 
obtained from a semi-circular spectral function $-\frac{1}{\pi} \mathcal{I} \Delta(\omega) = \frac{1}{2\pi} 
\sqrt{1-\omega^2}$. The Trotter time-step was $\dt = 0.05$. The truncated weight was set to $10^{-12}$ during DMRG as 
well as during time evolution and the bond dimensions were not restricted to any maximal value. }
 \label{fig:U0_ChainVSStar}
 \end{figure}

 \begin{figure}
  \centering
  \includegraphics{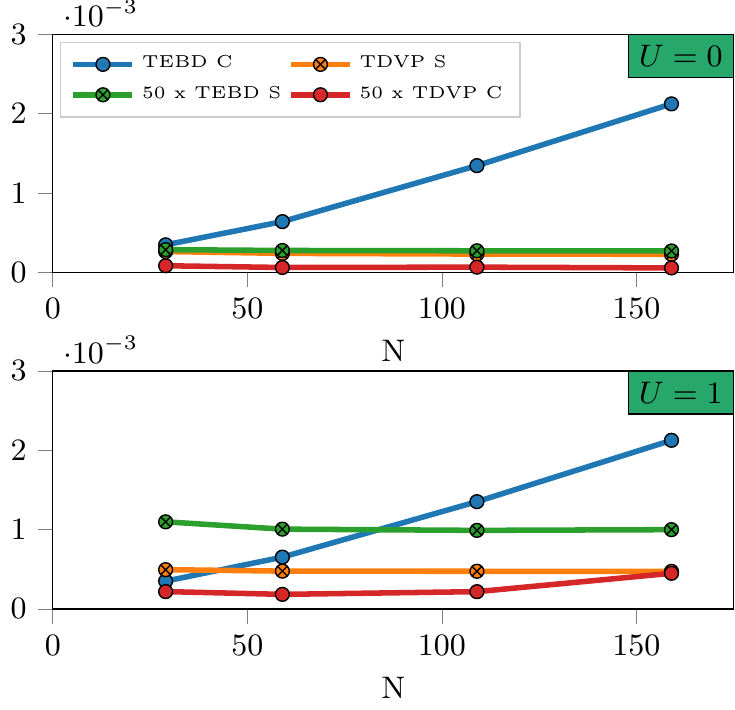}
  \caption{Maximum error for times $20<t<30$ as a function of bath size. Top: $U=0$ calculation shown in 
Fig.~\ref{fig:U0_ChainVSStar}. Bottom: $U=1$ calculation shown in Fig.~\ref{fig:U_ChainVSStar}. Note that the errors of 
TEBD-S and TDVP-C have been multiplied by a factor of 50 to make the graphs comparable. }
 \label{fig:ErrorVSBathSize}
 \end{figure}

To compare the different algorithms which have different sources of errors, we use the following strategy. We 
first fix the parameters (time step $\Delta t$ and truncated weight $t_w$) and compare the error of the Green's 
function without taking the 
computation time into account. In the next step, we then address the real question of computation time versus error.

\subsection{U=0}
Since interactions only affect the impurity degrees of freedom, it is reasonable to expect that most of the errors from the
approximations in the time evolution is already present in the non-interacting case. 
Starting with $U=0$ has the advantage of giving us access to the exact Green's function by diagonalization of the 
hopping matrix $T_{ij}$:
\begin{align}\label{eq:diag_Tij}
H = \sum_{ij} c^\dag_i T_{ij} c_j = \sum_{ijk} \underbrace{ c^\dag_i U^\dag_{ik}}_{c^\dag_k} E_k \underbrace{U_{kj}
c_j}_{c_k} = \sum_k E_k n_k ,
\end{align}
where $H= H_{\text{star}}$ or $H= H_{\text{chain}}$, respectively. Note that in the following, we compare each 
calculation to the exact Green's function for the system defined by its 
hopping matrix $T_{ij}$ for the finite size bath. 
\\
In Fig.~\ref{fig:U0_ChainVSStar} we compare the four time evolution schemes for various system sizes and for fixed 
parameters $\Delta t$ and truncation.
The truncated weight was chosen very small ($10^{-12}$), such that the main error source is the approximation of the 
time evolution operator, not the truncation of the tensor network. The magnitude of the error suggests that TDVP in 
chain geometry (TDVP-C) is the best algorithm followed by the TEBD in star geometry (TEBD-S). This does not consider 
the necessary bond dimension $m$ though, which was very different depending mostly on the bath geometry. While in star 
geometry the bond dimensions were $48$ and $62$ for TDVP and TEBD respectively, they grew to $145$ and $220$ in chain 
geometry. This difference has drastic effects on the computation time discussed below, which scales $\sim m^3$.
\\
Three observations are especially interesting in Fig.~\ref{fig:U0_ChainVSStar}. First, for TDVP the chain geometry 
gives more precise results, whereas for TEBD the star geometry gives a lower error. For TDVP this behavior may be related to
the projection error being zero in the chain geometry~\cite{HaegemanTDVP2}.

Second, in all cases, the error seems to be about constant in time. This is true until the particle added in the 
calculation of the Green's function is reflected at the boundaries of the finite size system (chain geometry) and then reaches the 
impurity again. Somewhat surprisingly, the same time scale also holds for 
the star geometry, even though the picture of a particle traveling towards the end of the bath applies only to the 
chain geometry. On the other hand the star geometry and the chain geometry are equivalent by a unitary transformation that can 
explain this time scale.
 The increase in error in the top plot of Fig.~\ref{fig:U0_ChainVSStar} (N=29) beginning at times $t>60$ 
is due to such a reflection.
\\
Third, as a function of bath size, only the error of TEBD-C shows the expected linear scaling with $N$ as also shown 
in Fig.~\ref{fig:ErrorVSBathSize} (top). The error of all other algorithms are almost exactly constant in the bath 
size. For TEBD-S and for TDVP, the error even seems to become slightly smaller as $N$ is increased. This is very 
surprising considering Eq.~\eqref{eq:TrotterErr_StarTot} and the $N$ successive approximations of TDVP.
\begin{figure}
 \begin{center}
  \includegraphics{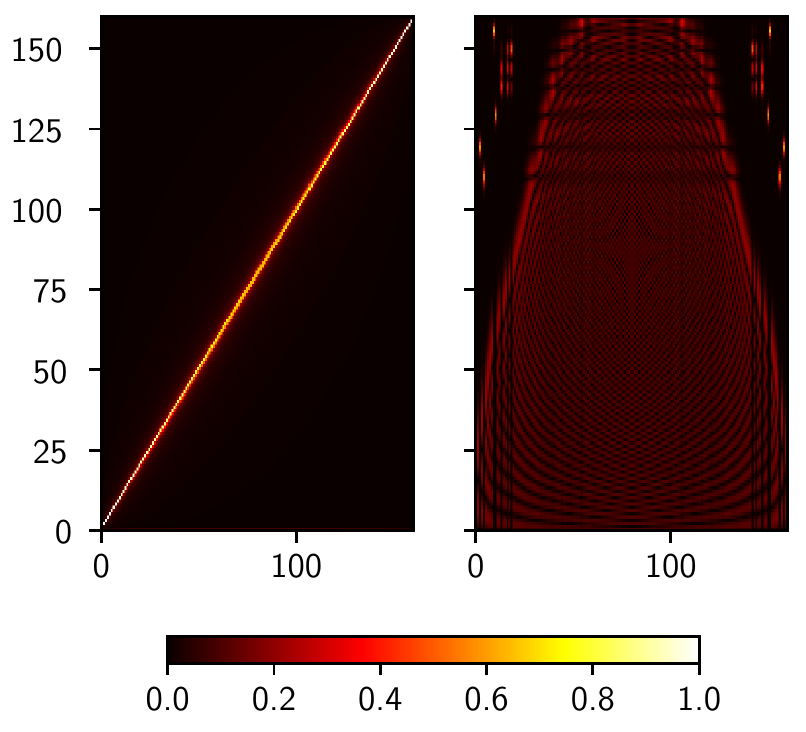}
  \end{center}
  \caption{Matrix elements $|U_{ik}|$ (see Eq.~\eqref{eq:diag_Tij}) in the star geometry (left) and in the chain 
geometry 
(right) for $N=159$ for the semi-circular bath as in Fig.~\ref{fig:U0_ChainVSStar}. While the non-zero entries in star 
geometry are strongly concentrated around the diagonal, $U_{ik}$ in the chain geometry has $\sim N^2$ relevant non-zero 
elements. Note that for the star geometry, the terms $U_{ik} = V_k$ for $i=I$ (see Eq.~\eqref{eq:Uik_star}) are located 
at the bottom of the plot, but 
are 
barely visible.}

  \label{fig:Uik}
\end{figure}

We can understand the observed behavior for TEBD in both geometries by examining the matrix elements of the Trotter 
error terms in Eq.~\eqref{eq:TrotterErr_StarTot} and Eq.~\eqref{eq:TrotterErrChain}. We start with the many-body ground 
state for $U=0$, 
i.e., the filled Fermi-sea (FS):
\begin{equation}
\ket{\psi_0} = \prod_{k \in FS} c^\dag_k \ket{0},
\end{equation}
where $k$ labels the eigenstates of the hopping matrix $T_{ij}$ (Eq.~\eqref{eq:diag_Tij}). As we have seen above, the 
leading errors of the Trotter decompositions correspond to hopping terms from one site to another. The error for the 
Green's function is then given by the matrix element of Eq.~\eqref{eq:TrotterErr_HbathStar} (star) and 
Eq.~\eqref{eq:TrotterErrChain} (chain) with the state $c_I^\dag \ket{\psi_0}$. 
The exact time evolution operator for $U=0$ is simply $U(t)=\prod_k e^{-i E_k t} n_k$, i.e., a time dependent phase 
factor for each occupied $k$. To find the magnitude and the number of terms contributing to the total error it hence suffices to evaluate (using the exact ground state (the Fermi-sea) for $\bra{\psi_0} $):
\begin{equation}\label{eq:ExpValue_hopp}
\bra{\psi_0} c_I c^\dag_i c_j c_I^\dag \ket{\psi_0}.
\end{equation}
The indices $i$ and $j$ are determined by the various hopping terms appearing in Eq.~\eqref{eq:TrotterErr_HbathStar} 
and 
Eq.~\eqref{eq:TrotterErrChain}. In $k$-space, we find:
\begin{align}\label{eq:ExpValue_hopp_kspace}
  \bra{\psi_0} c_I c^\dag_i c_j c_I^\dag \ket{\psi_0} = &\sum_{k>FS} U^\dag_{kI} U_{Ik} \sum_{k' \in FS} U^\dag_{k'i} 
U_{jk'} \\
+  &\sum_{k, k'>FS} U^\dag_{I k'} U_{kI} U^\dag_{ki} U_{jk'} \nonumber,
\end{align}
where $U_{ik}$ are the matrix elements of the unitary transformation in Eq.~\eqref{eq:diag_Tij}, and the index value 
$I$ 
again denotes the impurity.
Note that this expression is valid for both the star- and the chain geometry. Differences between them are encoded in 
the different unitary transformations $U_{ik}$ diagonalizing the hopping matrix $T_{ij}$. In star geometry, the bath 
states with energy $\epsilon_k$ are already very close to the eigenstates of $T_{ij}$. This implies that most entries 
in $U_{ik}$ are nearly zero, except for a few values around $i=k$. Indeed, as Fig.~\ref{fig:Uik} (left plot) demonstrates, 
$U_{ik}$  in star geometry has relevant non-zero entries only around the diagonal. This means that in star geometry 
there are only $\sim N$ relevant terms in the matrix products in Eq.~\ref{eq:ExpValue_hopp_kspace}. In other words, not all terms of 
$\mathcal{C}_{\text{star}}$ contribute and we expect an error independent of system size for the leading order $(\dt)^3$.
To make this point more clear, let us look at the case where only the diagonal of $U_{ik}$ contributes: 
\begin{equation}\label{eq:Uik_star}
  U_{ik}^{\text{star}} \approx  \left\{
    \begin{array}{lr}
      \delta_{i,k} \text{, for } i \neq I \\
      \sim V_k \text{, for } i = I
    \end{array}
  \right.
\end{equation}
We note that the approximation $U_{Ik} \sim V_k$ is in good agreement with the true form of $U_{ik}$.
Because of Eq.~\eqref{eq:Uik_star}, $\mathcal{C}_{\text{star}}$ does not scale with system size anymore, since 
$\delta_{ik}$ always removes at least one summation. This remains true, when $U_{ik}^{\text{star}}$ contains a finite 
width band of relevant values instead of just the diagonal. For the chain geometry on the other 
hand $U_{ik}$ is a full matrix with $\sim N^2$ entries of similar size, as shown in Fig.~\ref{fig:Uik} (right plot). 
Therefore, no such simplification occurs and the error should indeed scale linearly with $N$.
\\
For TDVP, such arguments do not hold, since its error is independent of the bath geometry used. The evaluation of the 
corresponding commutators is far from trivial and we hence leave this point open for future studies.

\subsection{Finite $U$}\label{sec:results_U}

\begin{figure}
  \centering
  \includegraphics{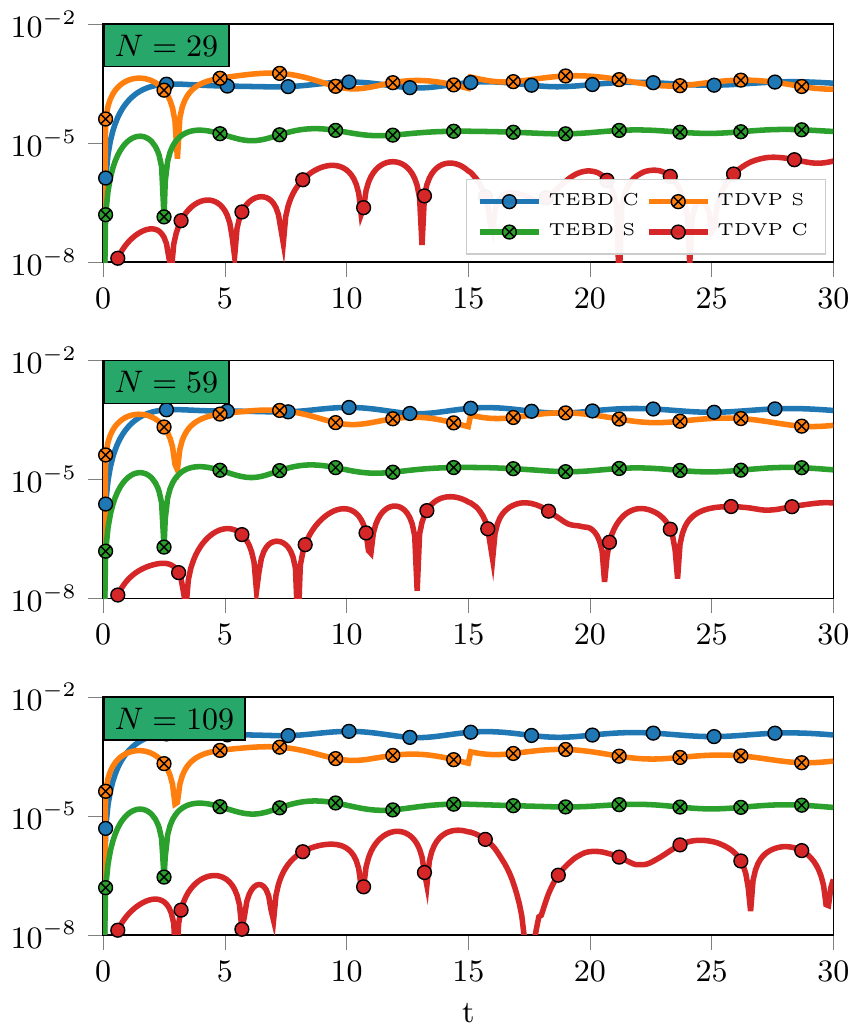}
  \caption{Absolute difference of $\Re G^{>}_{\text{ref}}(t)$ between high precision reference calculations and 
calculations with lower precision for various bath sizes and $U=1$. The bath was obtained from a semi-circular 
spectral function $-\frac{1}{\pi} \mathcal{I} \Delta(\omega) = \frac{1}{2\pi} \sqrt{1-\omega^2}$. \\
Because of the results at $U=0$, we chose TEBD for the star geometry and TDVP for the chain 
geometry for the reference calculations. These used a truncated weight of $10^{-14}$ ($10^{-13}$) and a Trotter 
time-step $\dt = 0.005$ ($\dt = 0.005$) for TEBD (TDVP) respectively. Parameters for the other calculations were 
$\dt = 0.05$ and truncated weight $10^{-12}$, the same as in Fig.\ref{fig:U0_ChainVSStar}. The bond dimensions of the MPS were not restricted to a maximal value. }
 \label{fig:U_ChainVSStar}
 \end{figure}

In Fig.~\ref{fig:U_ChainVSStar} we show a comparison similar to Fig.~\ref{fig:U0_ChainVSStar}, now for $U=1$. Since in the 
interacting case we do not have access to the exact solution, we compare to reference calculations with very high 
precision done separately for each bath size (see Fig.~\ref{fig:U_ChainVSStar} for details). Overall, we find that the 
errors of lower precision calculations in 
Fig.~\ref{fig:U_ChainVSStar} are almost identical to the ones obtained in 
the non-interacting case. In particular, we again find that only the error of TEBD-C scales appreciably with system 
size (see Fig.~\ref{fig:ErrorVSBathSize}). 
\\
The necessary bond dimension on the other hand changes drastically. Using the semi-circular bath spectral function, the 
maximal value was $m_{\text{star} } = 364$ $(128)$ compared to
$m_{\text{chain} } = 260$ $(214)$ for TEBD (TDVP) respectively.
\\
So far, we only compared the error for a given set of parameters (i.e., truncation $t_w$ and time step $\Delta t$). The 
actual quantity of interest is the computation time for given accuracy (or vice versa). Although all algorithms scale$~\sim m^3$, 
their computation times are very different and comparisons are not straightforward for several reasons. 

First, in star geometry the bond
dimensions is strongly peaked around the center bath site, whereas in chain geometry it is more flat. Therefore, the 
maximal bond dimension is not a good indicator of actual computation times. Second, a single TDVP step is generally 
much more expensive than a TEBD step. For example, in the calculations shown in Fig.~\ref{fig:U_ChainVSStar}, TEBD-S is 
faster than TDVP-S although the maximal bond dimension is larger by a factor of three in TEBD-S. Third, TDVP 
generally allows for much larger time steps $\Delta t$ for a give accuracy. 

Additionally, the advantage of the different geometries will likely depend on the actual bath parameters. 
For DMFT, most of the calculations are performed close to the self consistent point. We therefore studied the 
computation times necessary to reach a prescribed precision at the self consistent point of the Bethe lattice at $U=2$ 
($U=4$ in the units of Ref.~\onlinecite{WolfStarG}) for a time evolution up to $t=15$.
\begingroup
\begin{table} 
\setlength\extrarowheight{3pt}
\setlength\extrarowheight{3pt}
\begin{tabular}{|c|cc|c c|} 
\hline
 \textbf{Alg.} &  \textbf{$\Delta t$} & \textbf{$t_w$} & \textbf{Error ($10^{-4}$)} & \textbf{Wall time (s)}\\ \hline

TEBD-C &  $0.005$ & $10^{-10}$  & $2.5$ & 2370   \\ \hline
TDVP-S &  $0.005$ & $10^{-12}$  & $1.0$ & 4434 \\ \hline
TEBD-S & 0.1  & $10^{-9}$  & 1.7  & 35   \\
TEBD-S & 0.01 & $10^{-11}$ & 0.14 & 384  \\
TEBD-S & 0.01 & $10^{-13}$ & 0.01 & 1655 \\ \hline
TDVP-C & 0.5  & $10^{-8}$  & 2.6  & 457  \\
TDVP-C & 0.1  & $10^{-10}$ & 0.2  & 1856 \\
TDVP-C & 0.05 & $10^{-12}$ & 0.02 & 7130 \\ \hline
\end{tabular}
\caption{Comparison of computation times for the $U=2$ Bethe lattice self consistent bath. We chose example parameters 
for the time step $\Delta t$ and the truncated weight $t_w$ to approximately obtain errors of different orders of 
magnitude $10^{-4}$ to $10^{-6}$. We used the largest time step out of $\{ 0.005, 
0.01,0.05,0.1,0.5 \} $ for which this error can be achieved with low enough truncated weights and then used the largest truncated weight at this time 
step with a similar error. For TEBD-C and TDVP-S only an error of about $10^{-4}$ was possible with the parameters studied. Reference calculations 
were again performed with TDVP-C in chain-geometry ($\Delta t = 0.005$, $t_w=10^{-13}$) and TEBD-S in star geometry
($\Delta t = 0.005$, $t_w=10^{-14}$). Wall times are reported for a Intel(R) Core(TM) i7-7740X CPU using a single thread and for a single calculation of $G^>(t)$ up to $t=15$ including DMRG for $N=59$. }
 \label{tab:comp_compTimes}
\end{table}
\endgroup
\\
Results are shown in Tab.~\ref{tab:comp_compTimes}. The two algorithms with large errors in Figs.~\ref{fig:U0_ChainVSStar} and \ref{fig:U_ChainVSStar}, TEBD-C and TDVP-S are not able to obtain errors smaller than approximately 
$10^{-4}$ for the parameters studied and are comparable in computation time (TDVP-S has about half the error of TEBD-C 
but double the computation time). Additionally, they are slow compared to the other two approaches confirming behavior seen in Fig.~\ref{fig:U0_ChainVSStar} and Fig.~\ref{fig:U_ChainVSStar}. 
\\
Let us now compare the two favorable algorithms, TDVP-C and TEBD-S. While from Fig.~\ref{fig:U0_ChainVSStar} and Fig.~\ref{fig:U_ChainVSStar} on would expect TDVP-C to be the better 
algorithm, Tab.~\ref{tab:comp_compTimes} clearly shows that TEBD in star geometry is actually superior to TDVP-C. 
Its computation times are lower by about a factor of $5$ or more for the same error. Conversely, for similar computation 
times, the error of TEBD-S is about one order of magnitude smaller than TDVP-C. 
This shows that TEBD in star geometry is the best approach to calculate impurity Green's functions and should also be 
considered the time evolution algorithm of choice for general non-equilibrium impurity problems as clever arrangement of sites allows time evolutions to be performed up to surprisingly long times~\cite{RamsNonEqTNS}.

\section{Conclusions} \label{sec:conclusions}
We compared tensor network time evolution algorithms for Anderson Impurity Models for different bath representations. 
Specifically, we used TEBD and TDVP for the star- as well as the chain-geometry. TDVP is readily applicable for 
the long-range hybridizations present in the star geometry. For TEBD this is not the case and we explained in some 
detail the adapted TEBD approach using swap-gates first published in Ref.~\onlinecite{FTPS}. Its major advantage is that each actual 
time evolution gate can be combined with a swap gate, involving no additional computational cost and thus preserving 
the simplicity of TEBD. 
For TEBD, we additionally performed an analytic calculation of the leading order of the error due to the Suzuki-Trotter 
decomposition in both the star- and chain-geometry. This and the approximations of TDVP led us to expect that the error 
in the Green's function should be proportional to the system size $N$ in all four cases. 
Surprisingly, we found that only TEBD in chain-geometry shows this behavior.
We were able to find an analytical explanation of the better scaling of TEBD in star geometry from the fact that the bath states in star geometry 
are already a good approximation to the single particle eigenbasis and therefore, most error terms do not contribute. 
For DMFT calculations, such a favorable scaling with system size is especially important, since one has to make sure to 
use large bath sizes $N$ to represent the bath hybridization well enough to reach 
the correct self-consistent point. 
\\
Regarding the magnitude of the error, it is important to use the best combination of time evolution algorithm and bath 
representation. TDVP is more precise in the chain geometry likely due to the absence of projection error in the chain geometry. For TEBD it turned out to be vice versa, i.e., star geometry has a lower 
error. With given set of parameters (time step $\Delta t$ and truncated weight $t_w$) TDVP in chain geometry has the 
lowest error. On the other hand, the actual quantity of interest is the computation time for a given maximal error. With 
this metric, we found that TEBD in star geometry is the most favorable algorithm, faster than TDVP in chain 
geometry by about a factor of $5$ or more. Combined with its general simplicity and stability, this makes TEBD in star geometry 
at present the best approach to solve impurity problems using real-time evolution.

\bibliography{Citations.bib}
\end{document}